\documentclass[smallextended]{svjour3}       
\smartqed  
\usepackage{graphicx}

\begin{document}

\title{An integro-differential equation for electromagnetic fields in linear dispersive media}


\author{V.A. Coelho \and F.S.S. Rosa \and Reinaldo de Melo e Souza \and C. Farina \and M.V. Cougo-Pinto     
}

\institute{V.A. Coelho, F.S.S. Rosa, C. Farina and M.V. Cougo-Pinto \at
              Instituto de F\'\i sica, UFRJ, CP 68528, Rio de Janeiro, RJ 21941-972 \\
                \email{farina@if.ufrj.br}           
                           \and
           Reinaldo de Melo e Souza \at
            Instituto de F\'\i sica, UFF, Niter\'oi, RJ 24210-346 \\
            \email{reinaldos@id.uff.br}
}

\date{Received: date / Accepted: date}

\maketitle

\begin{abstract}
We extend the usual derivation of the wave equation from Maxwell's equations in vacuum to the case of electromagnetic fields in dispersive homo\-geneous isotropic linear media.  Usually,  dispersive properties of materials are studied in Fourier space. However, it can be rewarding to consider these properties as causal functions of time.  Due to temporal non locality, this procedure gives rise to an integro-differential equation for the electromagnetic fields, that we also call a wave equation. We have not found this equation in the literature and we show in this paper why it can be useful.    
\end{abstract}

\section{Introduction}
\label{intro}

The behavior of electromagnetic fields in material media is not always intuitive and depends substantially on the particular properties of each medium. This is due to the fact that every real medium presents dispersion, both in time and in space, and the amount of dispersion can drastically change how a medium responds to electromagnetic fields.  Throughout history, dispersion has fascinated many scientists and it was carefully studied by giants like Newton, Cauchy, Helmholtz, Lorentz, Sommerfeld and Brillouin (\cite{Darrigol2003,Sommerfeld1914,Brillouin2013}). In fact, Lorentz's atomistic approach and Sommerfeld's mathematical treatments for wave propagation in dispersive media still constitute the basis for modern approaches (\cite{Jackson1999}). In spite of its long history, many important aspects of dispersive media are still at the forefront of research. Amongst many contributions, we can mention the analysis of, the angular momentum of optical fields and the electromagnetic helicity in these media (\cite{Nori2017,Nori2018}), the investigation of  the very definition of light momentum (in the context of the hundred year old Abraham-Minkowski controversy, see \cite{Silveirinha2017} and references therein), and the new results on polariton excitations in dispersive media (\cite{Liberato2016,Partanen2017}).  Since Sommerfeld, most treatments of electromagnetic wave propagation are given in Fourier space, in contrast to what is usually done in (non-dispersive) free space.  The advantages of formulating the problem in Fourier space are twofold: (i) For most materials and not very intense fields, the constitutive equations are linear and (ii) the dispersion relation relates frequency and wavector for each Fourier component of the field. However, there are also some shortcomings.  Often, one needs the electromagnetic field in real space and in order to obtain it from its Fourier components it is necessary to perform an integration in frequency domain. The integrand involves the index of refraction $n(\omega)=\sqrt{\epsilon(\omega)\mu(\omega)}$, which is responsible for branch-points and branch-cuts in the integrand, making a direct evaluation of the integral a difficult task(\cite{Brillouin2013}). There are several techniques to avoid this problem, as for instance the modal expansion which has received a lot of attention recently (\cite{Gralak2018,Lalanne2018,Demesy2018}).

 Another possibility is to work directly in the real space. Furthermore, it is conceptually interesting to develop a formalism which involves only the real valued electromagnetic fields $\mathbf{E}(\mathbf{r},t)$ and $\mathbf{H}(\mathbf{r},t)$ instead of the complex auxiliary Fourier modes. In this letter we develop one such method, and show that instead of the usual wave equation we obtain in free space we are led to an integro-differential equation. That dispersion leads to integro-differential equations is well-known for general ondulatory phenomena (\cite{Rok2000,Weckner2007,Reddy2015,Yoshimura2016}) and it has also been demonstrated in some particular instances for electromagnetism (\cite{Gross1949,Halstuch2016,Lenhof2004}). 
 
 In this work, we generalize these results and show that an integro-differen\-tial equation follows directly from Maxwell's equations. We emphasize that such a result consists of a unique wave equation that governs the dynamics of all possible electromagnetic fields in linear dispersive media (in the frequency domain there is a different equation for each Fourier component), potentially leading to a deeper understanding of the dispersive properties of material media. At the end we discuss some particular examples.  

\section{Integro-differential wave equation}
\label{sec:1}

Time dispersion is directly related to temporal non-locality, i.e., the fact that a medium does not respond instantaneously to a given electromagnetic field. In this sense, for instance, the polarization ${\bf P}({\bf r},t)$ depends not only on the electric field at instant $t$ but also on earlier instants. Analogously, spatial dispersion means that the polarization ${\bf P}({\bf r},t)$ is spatially non-local, i.e., depends not only on the electric field at position ${\bf r}$ but also at neighboring points. The constitutive equation relating the vector ${\bf D}({\bf r},t) = \epsilon_0{\bf E}({\bf r},t) + {\bf P}({\bf r},t)$ (we are using SI units) and the electric field ${\bf E}({\bf r},t)$ for an isotropic linear dispersive medium is given by
 \begin{equation}
 {\bf D}({\bf r},t) = \int  dt^\prime \! \int d^3{\bf r}^\prime  g({\bf r},{\bf r}^\prime; t, t^\prime)\,{\bf E}({\bf r}^\prime,t^\prime) \, ,
 \end{equation}
 where the electric response function $g$ is a real function, which is non-local in both space and time due to dispersion. By assuming spatial and temporal homogeneity, we constrain the dependence of the function $g$ to the differences ${\bf r} - {\bf r}^\prime$ and $t-t'$, and therefore
 \begin{equation}
 {\bf D}({\bf r},t) = \int  dt^\prime \! \int d^3{\bf r}^\prime  g({\bf r} - {\bf r}^\prime; t - t^\prime)\,{\bf E}({\bf r}^\prime,t^\prime) \, ,
 \end{equation}
For the sake of simplicity, we additionally assume that the material media considered in this work do not exhibit spatial dispersion (in fact, for many purposes spatial dispersion can be neglected in dielectric materials). In this case, the response function can be written as $g({\bf r} - {\bf r}^\prime; t - t^\prime) = \delta({\bf r} - {\bf r}^\prime) \overline\epsilon(t - t^\prime)$, known as the spatial local limit, so that the previous equation reduce to (\cite{Zangwill2013,Landau1984})
 \begin{equation}\label{DE}
 {\bf D}({\bf r},t) = \int_{-\infty}^\infty \!\!dt^\prime\, \overline\epsilon(t - t^\prime)\,{\bf E}({\bf r},t^\prime) , 
   \end{equation}
 where $\overline{\epsilon}$ is a real function of $t-t^{\prime}$.
The overlined notation $\overline{\epsilon}$ is to avoid the use of $\epsilon$  commonly reserved in the literature to denote electric permittivity as a function of frequency. Furthermore, we impose a causality requirement for $\overline{\epsilon}$, meaning that it must vanish for $\tau:=t-t'<0$ (the effect cannot precede the cause), and also that $\overline{\epsilon}(\tau) \rightarrow 0$ for $\tau \rightarrow \infty$ (the remote past can be neglected)  (\cite{Zangwill2013,Landau1984}). Finally, it should be said that an analogous equation relates the magnetic induction  
 ${\bf B}({\bf r},t) = \mu_0\left[{\bf H}({\bf r},t) + {\bf M}({\bf r},t)\right]$, where ${\bf M}({\bf r},t)$ is the magnetization,  to the magnetic field ${\bf H}({\bf r},t)$, namely,
 \begin{equation}\label{BH}
  {\bf B}({\bf r},t) = \int_{-\infty}^\infty \!\! dt^\prime \,\overline\mu(t - t^\prime)\,{\bf H}({\bf r},t^\prime) \; , 
   \end{equation} 
where the magnetic response function $\overline{\mu}$  is also a real function that satisfies conditions analogous to those satisfied by the electric response function $\overline{\epsilon}$.
  
The past interval of time in which the electromagnetic fields have influence on a specific medium is of the order of its relaxation time \cite{Landau1984}.

Notice that in vacuum we do not have dispersion and thus $\overline{\epsilon}(t-t^{\prime})=\epsilon_0\delta(t-t^{\prime})$ and $\overline{\mu}(t-t^{\prime})=\mu_0\delta(t-t^{\prime})$, so that the constitutive relations (\ref{DE}) and (\ref{BH}) reduce to the vacuum constitutive relations. 

In order to derive the wave equation for electromagnetic fields in a temporal dispersive medium, we start by writing Maxwell's equations for the electromagnetic fields ${\bf E}$, ${\bf B}$, ${\bf D}$ and ${\bf H}$ in the absence of free charges and currents in such a medium, namely,
\begin{eqnarray}
\nabla\cdot{\bf B}({\bf r},t)=0\, \;\; &,& \nabla\times{\bf E}({\bf r},t)=-\frac{\partial{\bf B}(\mathbf{r},t)}{\partial t}  \label{Maxwellhomequationsinmedium}\\
\nabla\cdot{\bf D}({\bf r},t)=0 \, \;\; &,& \nabla\times{\bf H}({\bf r},t)=\frac{\partial{\bf D}({\bf r},t)}{\partial t}\, . \label{Maxwellinhomequationsinmedium}
\end{eqnarray}
These equations must be complemented by the two constitutive relations written in Eq(s) (\ref{DE}) and (\ref{BH}).  
 To obtain the desired wave equation in dispersive media we follow essentially the same  procedure 
used to obtain the wave equation in vacuum.  A first thing to notice is that since $\nabla\cdot{\bf D}({\bf r},t)=0$ then, from Eq.(\ref{DE}), we obtain\footnote{Note that we used causality in order to rewrite the limits of integration} $\int_{-\infty}^t \!\!dt^\prime\, \overline\epsilon(t - t^\prime)\,\nabla\cdot{\bf E}({\bf r},t^\prime)=0$. Since this must be valid for any $t$ we see that $\nabla\cdot{\bf E}({\bf r},t)=0$. 
In a completely analogous way it can be shown that  $\nabla\cdot{\bf B}({\bf r},t)={\bf 0}$ implies $\nabla\cdot{\bf H}({\bf r},t)={\bf 0}$. 
Taking the curl of both sides of the second equation in (\ref{Maxwellhomequationsinmedium}) and using the constitutive equation (\ref{BH}), we obtain
\begin{eqnarray}
\label{equacaodaondanomeioE1}
\nabla^2{\bf E}({\bf r},t) &=& \frac{\partial}{\partial t}\nabla\times
\int_{-\infty}^{\infty}dt^{\prime}\overline{\mu}(t-t^{\prime}){\bf H}({\bf r},t^{\prime})\cr
&=&
\int_{-\infty}^{\infty}dt^{\prime}     \;     \frac{\partial\overline{\mu}(t-t^{\prime})}{\partial t}
\frac{\partial{\bf D}({\bf r},t^\prime)}{\partial t^\prime}\cr
&=&
\int_{-\infty}^{\infty}dt^{\prime}\int_{-\infty}^{\infty}dt^{\prime\prime}
\,\frac{\partial\overline{\mu}(t-t^{\prime})}{\partial t}
\,\frac{\partial\overline{\epsilon}(t^{\prime}-t^{\prime\prime})}{\partial t^{\prime}}
\,{\bf E}({\bf r},t^{\prime\prime})\;,
\end{eqnarray}
where in the last step we used  the constitutive equation (\ref{DE}). 
Switching the order of integration in this equation and using the identity 
$\partial\overline{\mu}(t-t^{\prime})/\partial t=-\partial\overline{\mu}(t-t^{\prime})/\partial t^{\prime}$, we get
\begin{eqnarray}\label{equacaodaondanomeioE6}
\nabla^2{\bf E}({\bf r},t) &=& - \int_{-\infty}^{\infty}dt^{\prime\prime}\left[\int_{-\infty}^{\infty}dt^{\prime}
\,\frac{\partial\overline{\mu}(t-t^{\prime})}{\partial t^{\prime}}
\,\frac{\partial\overline{\epsilon}(t^{\prime}-t^{\prime\prime})}{\partial t^{\prime}}\right]
{\bf E}({\bf r},t^{\prime\prime})\cr
&=&
\int_{-\infty}^{\infty}dt^{\prime\prime}
\int_{-\infty}^{\infty}dt^{\prime}
\,\overline{\mu}(t-t^{\prime})
\,\frac{\partial^2\overline{\epsilon}(t^{\prime}-t^{\prime\prime})}{\partial t^{\prime\prime 2}}
\,{\bf E}({\bf r},t^{\prime\prime})\;.
\end{eqnarray}
where in the last step we made an integration by parts\footnote{Note that there are no surface terms, precisely due to our assumptions regarding $\epsilon(t-t')$.} in the variable $t^{\prime}$ and used the fact that $\partial\overline{\epsilon}(t'-t'')/ \partial t'=-\partial\overline{\epsilon}(t'-t'')/ \partial t''$.  Integrating by parts twice more, we finally obtain
\begin{equation}\label{equacaodaondanomeioE}
\nabla^2{\bf E}({\bf r},t)
-\int_{-\infty}^{\infty}dt^{\prime\prime}
\left[
\int_{-\infty}^{\infty}dt^{\prime}\overline{\mu}(t-t^{\prime})\overline{\epsilon}(t^{\prime}-t^{\prime\prime})
\right]
\frac{\partial^2{\bf E}({\bf r},t^{\prime\prime})}{\partial t^{\prime\prime\,2}}={\bf 0}\,.
\end{equation}
By an entirely analogous method we obtain for the magnetic field
\begin{equation}\label{equacaodaondanomeioH}
\nabla^2{\bf H}({\bf r},t)
-\int_{-\infty}^{\infty}dt^{\prime\prime}
\left[
\int_{-\infty}^{\infty}dt^{\prime}\overline{\epsilon}(t-t^{\prime})\overline{\mu}(t^{\prime}-t^{\prime\prime})
\right]
\frac{\partial^2{\bf H}({\bf r},t^{\prime\prime})}{\partial t^{\prime\prime\,2}}={\bf 0}\,.
\end{equation}

\section{Discussion}

Equations (\ref{equacaodaondanomeioE}) and (\ref{equacaodaondanomeioH}) are the main results of this work. 
After consulting a vast list of graduate and undergraduate textbooks(\cite{Jackson1999},\cite{Griffiths2017}-\cite{Kong1990}),  we were unable to find in the literature the above integrodifferential equation.
They generalize the wave equation for the electromagnetic fields in vacuum to the case where the fields propagate in an isotropic homogenous linear dispersive medium (with no spatial dispersion). The price to be paid is that now instead of a differential equation we have to deal with an integrodifferential equation. A few comments about these equations are in order here: 
 {\it (i)} In first place,  as a selfconsistency check, the above integrodifferential equations reduce to the usual differential wave equations in vacuum  in the local limit where $\overline{\epsilon}(t-t^{\prime}) \rightarrow \epsilon_0\delta(t-t^{\prime})$ and $\overline{\mu}(t-t^{\prime}) \rightarrow \mu_0\delta(t-t^{\prime})$, as expected. {\it (ii)} Note that wave equations  (\ref{equacaodaondanomeioE}) and (\ref{equacaodaondanomeioH}) are linear in the electromagnetic fields so that  linear superpositions of their solutions are also  solutions of these equations. At this point, it is important to emphasize that the Fourier components of the fields, $\mathbf{\mathcal{E}}(\mathbf{r},\omega)=\int_{-\infty}^{\infty}{\bf E}(\mathbf{r},t)e^{i\omega t}dt$, satisfy the well-known Helmholtz equation
\begin{equation}\label{Helmholtzequation}
\nabla^2{\mathbf{\mathcal{E}}}({\bf r})+\omega^2{\mu}(\omega){\epsilon}(\omega){\mathbf{\mathcal{E}}}({\bf r})={\bf 0}\, ,
\end{equation}
as can be straightforwardly verified. In the above expression $\epsilon(\omega)$ and $\mu(\omega)$ are the Fourier transforms of $\overline{\epsilon}(t-t')$ and $\overline{\mu}(t-t^\prime)$ respectively.  An analogous result holds for the Fourier components of the magnetic field.
In this way, the infinite set of differential (Helmholtz) equations, one for each frequency, are encoded in a single integrodifferential wave equation. Actually, the integrodifferential equations  (\ref{equacaodaondanomeioE}) and (\ref{equacaodaondanomeioH})  are satisfied by any electromagnetic fields which are solutions of Maxwell's equations (\ref{Maxwellhomequationsinmedium})-(\ref{Maxwellinhomequationsinmedium})  in isotropic homogeneous linear dispersive media without free charges and currents. 

One important situation is when dispersion effects  are weak, i.e., the response functions are very narrow in time. For the sake of simplicity let us assume a non-magnetic material, that is $\overline{\mu}(t-t')=\mu_0\delta(t-t')$ and suppose  the timescale for a significative variation of the electric field is much greater than the characteristic timescales of the the material. In these cases, we can perform a Taylor expansion in the electric field in Eq.(\ref{equacaodaondanomeioE}), namely, 
\begin{equation}
\frac{\partial^2{\bf E}({\bf r},t^\prime)}{\partial {t^\prime}^2} = 
\frac{\partial^2{\bf E}({\bf r},t)}{\partial t^2} + \frac{\partial^3{\bf E}({\bf r},t)}{\partial t^3}(t^\prime - t) + {\cal O}\Bigl((t^\prime - t)^2\Bigr)\, ,
\end{equation}
that once inserted into Eq.(\ref{equacaodaondanomeioE}) yields
\begin{eqnarray}
\nabla^2{\bf E}({\bf r},t) &-& \left[\mu_0 \int_{-\infty}^{\infty}dt^{\prime}\overline{\epsilon}(t-t^{\prime})\right]\frac{\partial^2{\bf E}({\bf r},t)}{\partial t^2} \;\; + \cr\cr
&+&
 \left[\mu_0\int_{-\infty}^{\infty}dt^{\prime}\overline{\epsilon}(t-t^{\prime})(t^\prime - t)\right]\frac{\partial^3{\bf E}({\bf r},t)}{\partial t^3} \;=\; {\bf 0} \, ,
\end{eqnarray}
where we used that $\overline{\mu}(t-t')=\mu_0\delta(t-t')$.
This is a partial differential equation with a third-order time derivative. If we include more and more dispersive corrections, higher and higher order  time derivatives in the electromagnetic fields will show up. 
The above expression can be written in a more appealing form
\begin{eqnarray}
\nabla^2{\bf E}({\bf r},t) -\mu_0\epsilon(0)\frac{\partial^2{\bf E}({\bf r},t)}{\partial t^2} +i\mu_0\epsilon'(\omega)|_{\omega=0}\frac{\partial^3{\bf E}({\bf r},t)}{\partial t^3} \;=\; {\bf 0} \, , \label{3owaveeq}
\end{eqnarray}
where $\epsilon(0)$ corresponds to the zero frequency component of the Fourier transform of $\overline{\epsilon} (t-t')$ and $\epsilon'(\omega)=d\epsilon(\omega)/d\omega$. This procedure can be easily generalized to obtain higher order terms and it can be shown that all the coefficients in the above equation are real.

\section{Conclusion}

In this work we have studied the dynamics of electromagnetic fields in a dispersive homogeneous isotropic linear medium directly in the real space (instead of in the Fourier frequency space).  Since dispersion entails a time memory in the medium response functions, the electromagnetic fields satisfy an integro-differential equation which subsume the infinite set of Helmholtz differential equations for each Fourier component. We illustrated our equation in two examples, the monochromatic limit and the weak dispersive regime. In the latter, we showed that the effect of dispersion consists in introducing extra terms to the standard wave equation with higher-order time derivatives of the fields. We believe that this equation may provide new approaches in dealing with electromagnetic dispersion phenomena.

\begin{acknowledgements}
We are indebted to Prof(s) P. A. Maia Neto, C. E. Magalh\~aes de Aguiar and F.A. Pinheiro for helpful discussions. The authors also thank Prof. Richard Price for a constructive criticism on this work. The authors thank the Brazilian agencies National Council for Scientific and Technological Development (CNPq) and Carlos Chagas Filho Foundation for Research Support of Rio de Janeiro (FAPERJ) for support.
\end{acknowledgements}

\end{document}